\documentstyle[12pt]{article}

\newcommand{\EQ}{\begin{equation}}
\newcommand{\EN}{\end{equation}}
\newcommand{\bea}{\begin{eqnarray}}
\newcommand{\ena}{\end{eqnarray}}
\newcommand{\vs}[1]{\vspace{#1 mm}}

\renewcommand{\a}{\alpha}

\newcommand{\pa}{\partial}

\newcommand{\uda}{\nearrow \kern-1em \searrow}
\newcommand{\zp}{z_+}
\newcommand{\zm}{z_-}
\newcommand{\xp}{x_+}
\newcommand{\xm}{x_-}

\begin{document}

\topmargin 0pt
\oddsidemargin 5mm

\renewcommand{\Im}{{\rm Im}\,}
\newcommand{\NP}[1]{Nucl.\ Phys.\ {\bf #1}}
\newcommand{\PL}[1]{Phys.\ Lett.\ {\bf #1}}
\newcommand{\CMP}[1]{Comm.\ Math.\ Phys.\ {\bf #1}}
\newcommand{\PR}[1]{Phys.\ Rev.\ {\bf #1}}
\newcommand{\PRL}[1]{Phys.\ Rev.\ Lett.\ {\bf #1}}
\newcommand{\PTP}[1]{Prog.\ Theor.\ Phys.\ {\bf #1}}
\newcommand{\PTPS}[1]{Prog.\ Theor.\ Phys.\ Suppl.\ {\bf #1}}
\newcommand{\MPL}[1]{Mod.\ Phys.\ Lett.\ {\bf #1}}
\newcommand{\IJMP}[1]{Int.\ Jour.\ Mod.\ Phys.\ {\bf #1}}

\begin{titlepage}
\setcounter{page}{0}
\begin{flushright}
OU-HET 195  \\
\end{flushright}

\vs{15}
\begin{center}
{\Large CLASSICALLY INTEGRABLE COSMOLOGICAL MODELS \\
WITH A SCALAR FIELD}
\vs{15}

{\large  Hisao Suzuki, Eiichi Takasugi and Yasuhiro Takayama}\\
\vs{8}
{\em Department of Physics, Osaka University \\
 1-16 Toyonaka, Osaka 560, Japan} \\
\end{center}
\vs{10}

\centerline{{\bf{Abstract}}}

New classes of classically integrable models in
the cosmological theories with a scalar field
are obtained by using freedoms of defining time and fields.
 In particular, some models with the sum of
exponential potentials in the flat spatial metric
are shown to be integrable.  The model with
the Sine-Gordon potential can be solved in terms of
analytic continuation of the non-periodic Toda field theory.

\end{titlepage}
\newpage
\renewcommand{\thefootnote}{\arabic{footnote}}
\setcounter{footnote}{0}

After the original inflational scenario was presented, various models
have been proposed to obtain a natural inflation such as
the chaotic inflation and the modified Brans-Diche theories.
In all of these scenarios, scalar fields play an important role.
In other words, in the early universe, the dynamics
seems to be governed by the gravity theory coupled with
a scalar field.
\par
As for classical solutions, it is known that we can easily get
them once we start by assuming the time variation of the scale
factor $a$: The classical solution of a scalar field and the potential
are expressed as functions of $a$ and its derivatives so that
in principle the potential is expressed as a function of a scalar
field[1].   Moreover, equations for gauge invariant variables for
scalar perturbations can be reduced to a second
order homogenious differential equation whose coeficients are
solely determined by $a$ and its derivatives. By solving this
eauation, several analytic solutions for the inflation models
are derived[1].
\par
In this paper, we present a method to get classically integrable models
for gravitational theories coupled with a scalar field in the case
where the dilatational mode and scalar field do not have spatial
dependence. A method to get integrable models was proposed in Ref.[2],
but unfortunately the integrable models were very few.
Here, we show  that with a slight modification
of their approach,  we can get integrable models for the
larger class of models.  In particular, it will be shown that
models with the exponential potential corresponding to
the Brans-Dicke theory are classically integrable.
We also find the Sine-Gordon theory coupled to
gravity can be solved in terms of Toda theories.
\par
The action of the gravity coupled to a scalar field is given by
\EQ
S =  \int d^4x \sqrt{-g}[ { 1 \over {16 \pi G}} f(\Phi)R + { 1 \over2}
(T(\Phi) g^{\mu\nu}\partial_\mu \Phi \partial_\nu \Phi - U(\Phi))],
\EN
where $f(\Phi)$, $T(\Phi)$ and $U(\Phi)$
are functions of $\Phi$.
 The standard
choice of the action is $f=1 - \xi \Phi^2, T=1$ (the choice
$ \xi = { 1 \over 6}$ is called minimal coupling).
The modified Brans-Dicke theories
correspond to the choice $f=\Phi$, $T = \omega(\Phi) / \Phi^2$,
where
the ordinary Brans-Dicke theory is given
by $\omega = 1$, and other choice is
called the extended Brans-Dicke theories[3].
For $T(\Phi)=0$, the scalar field is just
the auxirialy field and by eliminating it with the use of
equation of motion, we get the action of the higher derivative
gravity[4].

It is known that the general form of the action in Eq.(1)
can be transformed to a
action in the Einstein frame as followings[1]:
We make a conformal transformation
\EQ
g_{\mu\nu} = { 1 \over f(\Phi)}{\tilde{g}}^{\mu\nu},
\EN
and  define a new scalar field as
\EQ
d\chi = [ T(\Phi) f^{-1} - 3 ({f' \over f})^2] d \Phi.
\EN
Then, we change the scalar potential as
\EQ
\tilde U(\chi) \equiv U(\Phi) f^4(\Phi)
\EN
and finally we get the action in the Einstein frame:
\EQ
S = \int d^4x \sqrt{-\tilde{g}} [ { 1 \over {16 \pi G}}
 R({\tilde{g}}) + { 1 \over 2} (\tilde{g}_{\mu\nu}
\partial_\mu \chi \partial_\nu \chi - { 1 \over 2} \tilde U(\chi))].
\EN
Except for the singular case
$ T f^{-1} - 3({f' \over f})^2 =0$ where
we just obtain the action with no kinetic term of the scalar field.
Hereafter, we consider the non singular case and thus we treat
the action in Eq.(1) with $T=f=1$.

We start from
the metric of the universe to be the spatially homogeneous metric as
\EQ
(ds)^2 =\sigma^2( N^2 (dt_N)^2 - a^2(t_N)[ { (dr)^2 \over 1 - Kr^2}
+ r^2 (d\theta)^2
+ r^2 \sin^2 \theta (d \varphi)^2]),
\EN
where $N$ is the gauge function and $\sigma^2=2G/3\pi$.
We assume that $\phi$ depends only on the time $t_N$. Writting
\EQ
\phi =(2\pi^2\sigma^2)^{1 \over 2} \Phi(t_N),  \qquad
V(\phi)=2\pi^2\sigma^4U(\Phi),
\EN
the $1$ dimensional action in Eq.(1) with $f=T=1$
is expressed as follows:
\EQ
S= { 1 \over 2} \int dt_N[ N^{-1}(- a \dot{a}^2 + a^3 \dot{\phi}^2) +
N( - a^3 V(\phi) + K a)],
\EN
where we have omitted the volume factor of the space.
Note that the variation with respect to $N$ gives
us the hamiltonian constraint
\EQ
 ({\dot{a} \over a})^2 - \dot{\phi}^2 - V  + {K\over a^2}= 0.
\EN
An important point in our method is that
the equation of motion does not depend on the choice of this
gauge function $N$. In other words, we can choose this function
in such a way that the system becomes easy to be solved.
This is the main advancement from the method discussed in Ref.[2].

We are now going to change the action in Eq.(8) to the one with ordinary
kinetic
terms.  All we have to do is to change variables $a$ and $\phi$ to
$z$ and $w$ in such a
way that the action is transformed to
\EQ
S= \int dt_N [(2 \tilde{N})^{-1} (- \dot{z}^2 + \dot{w}^2) - \tilde{N} W],
\EN
or by introducing variables $z_\pm = z \pm w$,
we write Eq.(10) as
\EQ
S= \int dt_N [-( 2 \tilde{N})^{-1} \dot{z}_+ \dot{z}_- - \tilde{N} W(z_+,z_-)].
\EN
When we set $\tilde{N} = 1$, the cosmological problem turns
out to be just
an ordinary kinematical problem with indefinite metric.

In order to find new variables, we define
\EQ
a = e^{h(z_+,z_-)},\qquad
\tau_\pm = h \pm \phi.
\EN
Then the action in Eq.(8) is written as
\EQ
S= { 1 \over2} \int dt_N
[ - { e^{{3 \over 2}(\tau_+ + \tau_-)} \over N}
\dot\tau_+ \dot\tau_- + N[- e^{{3 \over 2}(\tau_+ + \tau_-)}
V({ 1 \over2}(\tau_+ - \tau_-)) + K e^{{1 \over 2 }(\tau_+ + \tau_-)}].
\EN
The change of variables which preserves the form of kinetic term is know as
 the two dimensional conformal transformation with respect to the
fields $\tau_+$ and $\tau_-$. We therefore take the following choice
of variables:
\EQ
\tau_+ = \tau_+(z_+),\qquad
\tau_- = \tau_-(z_-).
\EN
By choosing gauge function as
\EQ
N = \tilde{N} e^{{ 3 \over 2}(\tau_+ + \tau_-)} \partial_+ \tau_+
\partial_- \tau_-,
\EN
we find that the kinetic terms of Eq.(13) transforms to those of (11),
as desired. Here $\partial_+$ and $\partial_-$ mean  derivatives with
respect to $z_+$ and $ z_-$, respectively.
To simplify  Eq.(15), we will use the following functions in stead of
$\tau_\pm$:
\EQ
x_+(z_+) = e^{{3 \over 2} \tau_+},\qquad
x_-(z_-) = e^{{3 \over 2} \tau_-}
\EN
Then Eq.(15) is just written as
\EQ
N = { 4 \over 9} \tilde{N} \partial_+ x_+ \partial_- x_-.
\EN
In terms of $x_+$ and $x_-$, the original fields $a$ and $\phi$ can be written
as
\EQ
a = (x_+ x_-)^{1 \over3}, \qquad
\phi = { 1 \over 3} \ln { x_+ \over x_-}.
\EN
We can also find that the potential term in Eq.(10) or (11) is given by
\EQ
W(\zp,\zm) =  { 2 \over 9} \pa_+ \xp \pa_-
\xm[ \xp \xm V[{1 \over 3}
\ln{\xp \over \xm}] - K (\xp\xm)^{1 \over 3}].
\EN
Note that the form of the potential $W$ depends on the function
$\xp$ and $\xm$ which are arbitrary functions of $z_+$ and $z_-$,
respectively, while the kinetic term takes the form of
${\dot z}_+{\dot z}_-$ as seen in Eq.(11).
We utilize the freedom of choosing functions of $x_+$ and $x_-$
to simplify the potential $W$ such that differential equations
are solved.

 Firstly, we consider the exponential potential with $K=0$:
\begin{flushleft}
$ [{\rm A}]:\quad V(\phi) = \lambda e^{-\a \phi} \quad
  (\alpha\not= \pm 6)$
\end{flushleft}

This case is equivalent to the Brans-Dicke theory with a constant
potential and is
known to possess a power law inflational solution
$a = a_0(t+c)^\beta, \beta = { 12 \over \a^2}$,
as a special solution.  The special case of $\alpha=\pm 2$ in this case is also
the special case of [H] whcih we discuss later. The case [H] is well studied
and
the   analytic solution is obtained[5]. Thus, $\alpha=\pm 2$ case was known to
be
soluble.
But for other cases, the analytic solutions are not known before.

Here we show that the general solution is
obtained in our method. We take the ansatz
\EQ
   x_+ = \zp^l \qquad \xm = \zm^m.
\EN
When we take $l$ and $m$ as
\EQ
l= { i \over 2 - {  \a \over 3}}, \qquad m = {j \over 2 +
{\a \over 3}}
\EN
with $i$ and $j$ are integers, then the potential $W$ is written as
\EQ
W(\zp,\zm) = { 2 \lambda \over 36 - \a^2} ij \zp^{i-1}\zm^{j-1}.
\EN
These are valid for $ \a \not=\pm 6 $. When the choice
of the pair $(i,j)$ is one of (1,1), (1,2), (2,1), (1,3), (2,2)
and (3,1), the theory can be solved by the separation of variables.
In particular, if we take $i=j=1$, the potential becomes constant.
With the gauge choice $\tilde N=1$, the equations of motion are written
by $\ddot{z_+} = \ddot{z_-} =0$ so that
\EQ
z_+=a_1t_N+a_2 \qquad z_-=b_1t_N+b_2.
\EN
The Hamiltonian constraint ${ {\dot{z}_+\dot{z}_-}\over 2}-W=0$
gives a restriction
\EQ
a_1b_1={{4\lambda}\over{36-\alpha^2}}.
\EN
The scale factor $a$ and the scalar field $\phi$ are obtained from
Eq.(18) as functions of $t_N$ through Eq.(23). From Eq.(17) with $\tilde N=1$,
the time $t_N$ is expressed as a function of $t$ as
\EQ
t= {{4}\over {36-\a^2}} \int (a_1t_N+a_2)^{l-1}(b_1t_N+b_2)^{m-1}dt_N,
\EN
where $l$ and $m$ are given in Eq.(21) with $i=j=1$.
In order to compare the power low behavior given before, we
have to perform the above integration. Here we consider
the large $t_N$ or $t$ limit of $a$ and $\phi$. From Eq.(25), we have
$t \propto t_N^{l+m-1}$. Then, $a =z_+^{l\over 3}z_-^{m\over 3} \propto
t_N^{(l+m)/3} \propto t^{12 \over \a^2}$ which reproduces the special
solution stated before.

With the ansatz in Eq.(20), some models with potentials consisting of
the sum of exponential forms are solvable. We found the following
cases with $K=0$ are solvable:
\begin{flushleft}
$[{\rm B}]: V(\phi) = \lambda_1 e^{ 2 \phi}
+ \lambda_2 e^{-2 \phi}$
\end{flushleft}
We find with the ansatz in Eq.(20)
\EQ
W ={1\over 8}( \lambda_1 \zp + \lambda_2 \zm ) \qquad
{\rm with} \qquad l=m={3\over4}.
\EN
This case is the special case of [H] explained later so that this model was
known to be  soluble[5].\par
\begin{flushleft}
$[{\rm C}]:V(\phi) = \lambda_1 e^{{5  \over 18}\phi}
+ \lambda_2 e^{ - { 6 \over 5}\phi}
$
\end{flushleft}
We find with the ansatz in Eq.(20)
\EQ
W ={25\over144}( \lambda_1 \zp + \lambda_2 \zm^2 )\qquad
{\rm with} \qquad l={5\over8},\quad
 m={5\over4}.
\EN
Note that a new soluble model is derived from a model by changing
$\phi$ to -$\phi$. In this case, $W$ is obtained by changing $z_+$ and $z_-$
and $l$ and $m$. Thus, the case
\par
\begin{flushleft}
$[{\rm C'}]:V(\phi) = \lambda_1
 e^{-{18 \over 5}\phi} + \lambda_2 e^{{6 \over 5}\phi}
$
\end{flushleft}
is also soluble and with the ansatz in Eq.(20) we find
\EQ
\quad W={25\over144}( \lambda_1 \zm + \lambda_2 \zp^2 )\qquad
{\rm with} \qquad l={5\over4}, \quad
 m={5\over8}.
 \EN
\begin{flushleft}
$[{\rm D}]:
V(\phi) = \lambda_1 e^{ 3 \phi}
+ \lambda_2 e^{-3 \phi} + \lambda_3
$
\end{flushleft}
We find with the ansatz in Eq.(20)
\EQ
W ={2\over 9}( \lambda_1 \zp^2 + \lambda_2 \zm^2 + \lambda_3 \zp \zm)
\qquad {\rm with} \qquad l=m=1.
\EN
Solutions for all of above cases can be
derived following the same manner demonstrated
for the exponential potential case.
The special cases of [D],
the case with $\lambda_1, \lambda_2 >0$ and
$\lambda_3=-2\sqrt{\lambda_1\lambda_2}$ and
the case with $\lambda_3=0$ are solved by Ritis et. al.[2].

As for the exponential potential in the case [A], $\alpha=\pm 6$
are excluded. Here we consider these cases:\par
\begin{flushleft}
$[{\rm E}]:V(\phi) = \lambda e^{-  6 \phi}$
\end{flushleft}

By taking the ansatz
\EQ
x_+ = e^{A\zp}, \qquad \xm = \zm^{1 \over 4},
\EN
we get
\EQ
W={1\over 18}A\lambda.
\EN
Thus, this case is solvable. The case with
$\alpha=-6$ can be solved by exchinging $x_+$ into $x_-$, i.e.,
by taking $x_+=z_+^{1\over 4}$ and $x_-=e^{Az_-}$.

Let us examine some other type of ansatz,
\EQ
\xp = e^{A\zp},\qquad \xm = e^{B\zm}.
\EN
There are several solvable cases under this ansatz.
One is the potential\par
\begin{flushleft}
$[{\rm F}]:V(\phi) = \lambda_1 e^{\a \phi}
+ \lambda_2 e^{\beta \phi}, \qquad \alpha \beta =36 \hskip 2mm (\a \not =
\beta)$
\end{flushleft}
In this case, with the choice
\EQ
 B={{2+{\a \over 3}}\over {2-{\a \over 3}}}A ,
\EN
the potential reduces to
\EQ
W={2\over9}AB
\left( \lambda_1\exp[2A(2+{\alpha \over 3})z]
  + \lambda_2\exp [\frac{12}{\alpha}A(2+{\alpha \over 3})w]\right).
\EN

Another nontrivial example which cannot be solved
by the separation of variable is the case of a Sine-Gordon potential
\begin{flushleft}
$[{\rm G}]:V(\phi) = \lambda \cos 6\sqrt{3} \phi $
\end{flushleft}
In this case, we take
\EQ
\xp = (288)^{1\over2} e^{-{1 \over 8 }\zp},\qquad
\xm = (288)^{1\over2} e^{-{1 \over 8 }\zm}.
\EN
Then, we get the following potential:
\EQ
W = \lambda e^{-{ 1 \over 2}z}\cos{{\sqrt{3} \over 2} w}.
\EN
In this case, we can find the following conserved currents:
\bea
H &=& - { 1 \over 2} p_z^2 + {1 \over2}p_w^2 +
\lambda e^{-{ 1 \over 2}z}\cos{{\sqrt{3} \over 2} w}= 0,\\
I &=& { 1 \over2} p_z^2p_w + { 1 \over6} p_w^3 +
{ \sqrt{3} \over2} \lambda p_z e^{-{ z \over2}}
\sin {\sqrt{3} \over2 }w + { \lambda \over 2} p_w e^{-{ z \over2}}
\cos {\sqrt{3} \over2}w,
\ena
where $p_z$ and $p_w$ are conjugate momentum of $z$ and $w$, respectively.
Since we found two conserved currents, the model is soluble.
The origin of the higher dimensional conserved current is understood
by noticing that
the system can be obtained by the analytic continuation of the
nonperiodic Toda theory[6]. Explicitly, the conserved quantities $H$ and $I$
are constructed from a Lax pair $(L,M)$ of the nonperiodic Toda theory:
That is $H \propto Tr L^3$ and $I \propto Tr L^3$ with the identification
$q_1=-(z+\frac i{\sqrt 3} \omega)/4$, $q_2=\frac i{2\sqrt 3} \omega$ and
$q_3=-(q_1+q_2)$ where $q_1$, $q_2$ and $q_3$ are defined in Ref.6.

Finally, we consider the integrable models with $K \ne 0$. This case
is in general hard to be solved. Except for a trivial example
$ V = \Lambda$, we can find the soluble model with the potential,
\begin{flushleft}
$[{\rm H}]:V(\phi) = \lambda_1 e^{ 2 \phi }
+ \lambda_2 e^{-{2 \phi }}\qquad (K\not= 0)$
\end{flushleft}
Similarly to the case [B], we take the ansatz
\EQ
x_+=z_+^{3\overA@4}, \qquad x_-=z_-^{3\over4},
\EN
we obtain
\bea
W={1\over8}(\lambda_1z_+ +\lambda_2z_- - K)
\ena
so that this case is soluble. This is a famous model studeid extensively[5] and
is known to be solble.

It may be worthwhile to note that our result depends heavily
on the fact that we introduced only one scalar field. In this case,
we can utilize the 2$d$ conformal transformation with respect to $a$ and
$\phi$ and we found several solvable models. If we have $n$ scalar fields
and wish to use similar simmetry,
we may need to consider $n+1$ dimensional conformal
transformation. Thus, the integrable models are quite restrictive
except for $n=1$ because the 2 dimensional conformal transformation
is known to be special. Note also that
the basic form of the soluble potential is of the exponential type.
These
are quite analogous to the integrability of $2$ dimensional gravity.

We have found some class of cosmological models can be solved
in terms of
the appropriate choice of the gauge
function and the conformal transformation.
But the integrable models are rare and most
of the physically interesting examples are not integrable.
For example, the potential which corresponds to
$R^2$ gravity [7] can be transformed to the
quartic potential by an appropriate choice of
the conformal transformation.   Of course,
since we
do not know the whole classification of integrable models even in the
ordinary kinematic problem, we have an opportunity to get other integrable
models.
We hope
larger class of interesting models can be found and analyzed in our
method.

Finally, we point out that  for some potentials given above, the Wheeler-De
Witt
equation can be solved analytically. This equation for the case [H] ( and thus
for
the case [A] with $\a=\pm 2$ and the case [B]) was solved and the analytical
solution is given[5]. By using our transformations, we can get the analytical
 solutions for the cases [A], [D], [E] and [F] which will be given in the
 forthcomming paper.


\newpage
\begin{thebibliography}{99}
\bibitem{KST}M. Kuwahara, H. Suzuki and E. Takasugi, \PR{D50} (1994) 661.
\bibitem{RITIS}R. de Ritis et. al., \PR{D42} (1990) 1091;
\PL{A149} (1990) 79.
\bibitem{LS}D. La and  P.J.  Steinhardt, \PRL{62} (1989) 376;

 P.J. Steinhardt and  F.S. Acetta, \PRL{64} (1990) 2740;

 J. Garcia-Bellido and M. Quiros, \PL{B 243} (1990) 45.
\bibitem{W}B. Whitt, \PL{B 145} (1984) 176.

\bibitem{GHM }L.J. Garay, J.J. Halliwell and G.A. Marugan, \PR{D43} (1991)
2572.

\bibitem{OP}M.A. Olshanetsky and A.M. Perelomov,
Phys. Rep. {[bf C71} No.5 (1981) 313.

\bibitem{SBB}D.S. Salopek, J.R. Bond and J.M. Bardeen, \PR{D40} (1989) 1753.
\end{thebibliography}
\end{document}